\documentclass[conference]{IEEEtran}
\IEEEoverridecommandlockouts
\usepackage{bm}
\usepackage{amsmath}
\usepackage{amssymb}
\usepackage[lined,boxed,linesnumbered,ruled,commentsnumbered]{algorithm2e}
\usepackage{amsmath,amsthm,amssymb,amsfonts}
\usepackage{caption}
\usepackage{cite}
\usepackage{graphicx}
\usepackage{subfig}
\usepackage{textcomp}
\usepackage[normalem]{ulem}
\usepackage{verbatim}
\usepackage{xcolor}
\usepackage{siunitx}
\usepackage{multirow}
\usepackage {makecell}
\usepackage{colortbl}
\usepackage[hidelinks]{hyperref}
\usepackage{cleveref}
\begin{document}
\captionsetup[figure]{labelfont={},labelformat={default},labelsep=period,name={Fig.}}
\definecolor{mygreen}{HTML}{BBC6C8}
\definecolor{c1}{HTML}{5BA199}
\definecolor{mygray}{gray}{.9}
\title{
Battery-Care Resource Allocation and Task Offloading in Multi-Agent Post-Disaster \\MEC Environment}
\author{\IEEEauthorblockN{Yiwei Tang\IEEEauthorrefmark{1}\IEEEauthorrefmark{2},
Hualong Huang\IEEEauthorrefmark{1},
Wenhan Zhan\IEEEauthorrefmark{1},
Geyong Min\IEEEauthorrefmark{3} ,
Zhekai Duan\IEEEauthorrefmark{4},
Yuchuan Lei\IEEEauthorrefmark{1}\IEEEauthorrefmark{5},
\thanks{Wenhan Zhan is the corresponding author.}}
\IEEEauthorblockA{\IEEEauthorrefmark{1}University of Electronic Science and Technology of China, 
Chengdu, 611731 China}
\IEEEauthorblockA{\IEEEauthorrefmark{2}University of Glasgow, Glasgow, G12 8QQ UK}
\IEEEauthorblockA{\IEEEauthorrefmark{3}University of Exeter, Exeter, EX4 4QF UK.
}
\IEEEauthorblockA{\IEEEauthorrefmark{4}University of Edinburgh, Edinburgh, EH8 9JU UK
}
\IEEEauthorblockA{\IEEEauthorrefmark{5}Sichuan Branch of China Telecom Group Co., Ltd., 610000 China}

Emails: ywtang1027@163.com, hlhuang\_uestc@163.com, zhanwenhan@uestc.edu.cn, \\G.Min@exeter.ac.uk,
zhekaiduan2312@gmail.com, 15308187794@189.cn
}
\maketitle

\begin{abstract}
Being an up-and-coming application scenario of mobile edge computing (MEC), the post-disaster rescue suffers multitudinous computing-intensive tasks but unstably guaranteed network connectivity. In rescue environments, quality of service (QoS), such as task execution delay, energy consumption and battery state of health (SoH), is of significant meaning. This paper studies a multi-user post-disaster MEC environment with unstable 5G communication, where device-to-device (D2D) link communication and dynamic voltage and frequency scaling (DVFS) are adopted to balance each user's requirement for task delay and energy consumption. A battery degradation evaluation approach to prolong battery lifetime is also presented. The distributed optimization problem is formulated into a mixed cooperative-competitive (MCC) multi-agent Markov decision process (MAMDP) and is tackled with recurrent multi-agent Proximal Policy Optimization (rMAPPO). Extensive simulations and comprehensive comparisons with other representative algorithms clearly demonstrate the effectiveness of the proposed rMAPPO-based offloading scheme. 
\end{abstract}


\section{Introduction}
Fettered by the capability of CPUs and the capacity of batteries, even evaluated mobile devices (MDs) still keep modern citizens from pursuing exponentially climbing requirements on the data transmission rate and quality of service (QoS) \cite{b1}. Inspired by network edge equipment's increasing computing and storage capacity, MEC is considered the next-generation network architecture for its proximity to end users. Within the radio access network, offloading latency and energy consumption could be greatly alleviated. A latent beneficiary of MEC is the emergency response scene, in which the remote cloud may not be connected. Moreover, in extreme scenarios such as post-disaster environments, the mighty base stations may be unavailable. Besides traditional cellular links, device-to-device (D2D) links that establish inter-device communication would bring significant advantages for emergency response, improving the QoS and system robustness \cite{b2}.

Aiming at enhancing user experience, it is always a sophisticated challenge to offload computation tasks in a mobile edge computing (MEC) environment, with regard to user preference, network connection, device capacity or availability and increasingly popular battery life extension \cite{b4}. In terms of user preference, according to \cite{b5}, distinguished service demands differentiate delays, and the more urgent task would consume more energy. For network states, dynamic network malfunctions such as device disconnection and transmission failure are always unavoidable in extreme emergencies even though technical advancement\cite{b2,b6}. Communication and computation ability are generally vital since MDs and edge devices (EDs) hold distinct aptitudes \cite{b3}.

With considerable spotlight, battery aging is regarded as one of the major obstacles for battery-supported devices. And up to scarce facilities in post-disaster scenes, battery lifetime extension is aspired desperately. Although many researchers have explored energy-efficient schedules \cite{b7,b8} and lower power is proven to prolong battery life \cite{b9}, battery degradation extent is barely kept an eye on, relieving only temporary pressure. Nevertheless, battery lifetime extension collides with the assignment delay ease, which calls for speedy running. To counterbalance these contradictory energy cost requirements, dynamic voltage and frequency scaling (DVFS) is proposed for flexibly-adjusted CPU operating frequency \cite{b10}.

The prime goal of computation offloading is to achieve the most favorable offloading decision and resource allocation. Although various approaches are available, few are appropriate, especially in multi-user circumstances \cite{b11}. Convex optimization, Lyapunov optimization, and deep reinforcement learning (DRL) are eminent in centralized conditions, although only the third one vanquishes one-step and scalability limitation. To deal with the centralized decision problem of DRL, game theory is devised for the distributed scenarios, despite its intense demand of prior knowledge. Therefore, multi-agent deep reinforcement learning (MADRL), a decentralized and self-learning method via agents' evaluation and interaction, becomes promising. However, on the large-scale post-disaster ruins, computation offloading and resource allocation might deviate from traditional fully cooperative optimization, instead being a mixed cooperative-competitive (MCC) situation for limited source and dramatic environment change sensitivity.

In this article, we first formulate a computation offloading and resource allocation optimization problem of a multi-user 5G post-disaster MEC environment supporting D2D communication and DVFS, in which a task would be executed locally, otherwise offloaded to other MDs or EDs. The objective is to enable each MD to individually develop a resource-efficient, delay-aware offloading model pandering to user preference and battery state of health (SoH). To adapt to such a dynamic problem, recurrent multi-agent Proximal Policy Optimization (rMAPPO), a MADRL algorithm powered by recurrent neural networks (RNNs), is proposed as a compensation of multi-agent Markov decision process (MAMDP), for its capture of long-term time dependency. We also embed distributed agents with separate rewards to actualize the MCC. Finally, extensive simulations involving comparison with representative algorithms are conducted. The experiment results validate the effectiveness of the proposed scheme, which arrives at a better equilibrium position.

\section{System Model}
\begin{figure}[htb]
    \vspace{-0.4cm}
    \centering\includegraphics[width=0.7\linewidth]{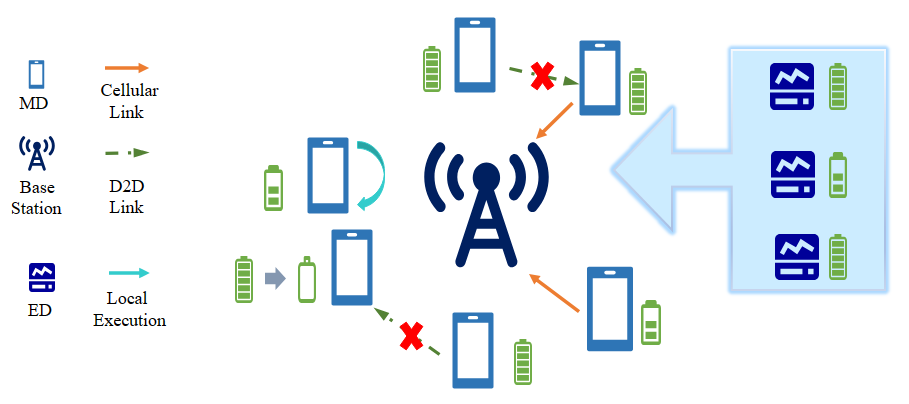}
    \caption{Overview of multi-user post-disaster MEC system.}
    \label{fig:MEC}
\end{figure}

In this paper, we consider a typical MEC post-disaster scenario, as shown in Fig. \ref{fig:MEC}. The system time is discretized into equal length slots, which is represented by $\mathbb{T}=\{t \mid t \in[1, T]\}$ and each slot lasts $\Delta t$. The important notations adopted in the paper are summarized in Table \ref{tab:Notation}.

\begin{table}[!htbp]
\vspace{-0.35cm}
\caption{Summary of Notations}
\centering
\begin{tabular}{ll}
\multicolumn{1}{c}{Notation}& \multicolumn{1}{c}{Definition} \\
\hline$F$ &\cellcolor{mygreen}{The signal-to-interference noise ratio (SINR)} \\
$R_{\text{tra}}$ &\cellcolor{mygray}{The transmission rate} \\
$\varepsilon_{\text{suc}}$ &\cellcolor{mygreen}{The SINR threshold of successful transmission } \\
$p_{\text{suc}}$ &\cellcolor{mygray}{The successful transmission probability} \\
$Ds$ &\cellcolor{mygreen}{The drop state} \\
$T_{\text{tra}}$ &\cellcolor{mygray}{The transmission time} \\
$E_{\text{tra}}$ &\cellcolor{mygreen}{The transmitter energy cost} \\
$E_{\text{rec}}$ &\cellcolor{mygray}{The receiver energy cost} \\
$T_{\text{fin}}$ &\cellcolor{mygreen}{The execution finish time} \\
$T_{\text{sta}}$ &\cellcolor{mygray}{The execution start time} \\
$T_{\text{exe}}$ &\cellcolor{mygreen}{The task execution time expense} \\
$E_{\text{exe}}$ &\cellcolor{mygray}{The task execution energy consumption} \\
$Fs$ &\cellcolor{mygreen}{The failure state} \\
$SoC_{\text{avg}}$ &\cellcolor{mygray}{The SoC average} \\
$SoC_{\text{dev}}$ &\cellcolor{mygreen}{The SoC standard deviation} \\
$N_{\text{cyc}}$ &\cellcolor{mygray}{The effective cycle number} \\
$BD$ &\cellcolor{mygreen}{The battery degradation extent} \\
$E_{\text{pri}}$ &\cellcolor{mygray}{The prioritized energy} \\
$BD_{\text{tot}}$ &\cellcolor{mygreen}{Total battery degradation extent of transmitter and receiver} \\
\hline
\end{tabular}
\label{tab:Notation}
\end{table}

\subsection{Device Model}
A set of MDs $\mathbb{M}=\left\{m^j \mid j \in[1, M]\right\}$ and a base station construct the MEC environment in which EDs $\mathbb{N}=\left\{n^i \mid i \in[1, N]\right\}$ are connected to the station via wired link, making negligible transmission delay \cite{b8}. Hence, practically working equipment at $t$th slot could be expressed as $Q=\left\{q^{h,t} \mid h \in\left[1, M+N\right]\right\}$. It needs to be noted that all devices, including EDs and MDs, are powered by rechargeable batteries and probably disconnect from each other because of hostile network quality or equipment breakdown after the disaster. With $\chi$ signing device disconnection, the battery level of the inaccessible device at slot $t$ would be observed as 0, leading to actual energy $\{B^{h,t}\mid B^{h,t}\in\left[0,B_{max}-\chi\left(B_{max})\right]\right\}$ within maximal battery energy $B_{max}$ and furthermore state of charge (SoC) $\{SoC^{h,t} \mid SoC^{h,t} \in\left[0,1]\right\}$. 

There is an individual agent for every MD. At the start of every slot, it monitors the environment before determining the offloading target and adjusting the objective's operating frequency through DVFS for its generated task. Afterward, the task is transmitted to the task queue of a scheduled target. 

\subsection{Task Model}
It is assumed that every MD produces its task with a specific possibility when a slot starts, inducing a discontinuous generation slot set $k \subseteq K, K \subseteq \mathbb{T}$. Therefore, if at slot $k$ MD $i$ generates its $g$th task, that task is defined as $G^{i,k, g}=\left\{S^{i,g}, C^{i,g}, D^{i,g}, X^{i,g}\right\}$ where $S^{i,g}, C^{i,g}, D^{i,g}, X^{i,g}$ symbolize data size, demanded CPU cycles, maximum tolerable delay and task type respectively. Based on preference to distinguished tasks, tasks are delineated into three kinds, namely $X^{i,g} \in \{1,2,3\}$, with dissimilar tolerable delay scope \cite{b5}: the first group mainly concentrates on real-time information transfer, including audio, video, medical service and so on. For the second type, user-interaction assignment like online games, delay necessities always rely on subscriber predication. Ultimately, for $X = 3$, namely non-real-time service, they usually are the least stringent about the deadline. Once the deadline is missed, the unprocessed part of a task would be managed in the largest affordable CPU capacity immediately to gratify subscribers.
 
\subsection{Execution Model}

\textbf{Task Transmission \& Drop State:} We postulate an orthogonal frequency division multiple access (OFDMA) communication manner, despite D2D and cellular link discrimination. Here, subchannels are divided evenly to transmitters choosing the same receivers, and their inter-interference is neglected as exclusive frequency spectrum allocation. Due to  communication ability bottleneck, D2D has a more limited coverage than cellular link, resulting in differently-ranged inter-equipment distance $Dt=\left\{Dt_{\text{m}}^{i, i^{\prime}, t}, Dt_{\text{e}}^{i, j, t} \right\}$. In addition, device capability also diversifies transmit power $P t=\left\{P t_{\text{m}}, P t_{\text{e}}\right\}$ and receive power $P r=\left\{P r_{\text{m}}, P r_{\text{e}}\right\}$. Note that subscript `$\text{m}$' flags D2D link and `$\text{e}$' stands for cellular link.

In such an emergency, a new light on transmission failure should be cast. Assuming a stable transmitting channel, signal-to-interference-plus-noise ratio (SINR) $F$ \cite{b3} and transmitting rate $R_{\text{tra}}$ between transmitter $h$ and receiver $h^{\prime}$ is written as:
\begin{align}
F^{h,g}=&\frac{P t^{h,g} (Dt^{h,h^{\prime},g})^{-\alpha}}{N_0}, \\
R_{\text{tra}}^{h,g}=&\frac{W^{h, h^{\prime}}n_{\text{sub}}^{h,g}}{n_{\text{tot}}}\log \left(1+F^{h,g}\right),    
\end{align}
where $\alpha$ is the path loss exponent. $N_0$ is the noise power spectral density, $W$ means spectrum bandwidth assigned to the base station or MD receivers while $n_{\text{tra}}^{h,g} \in \mathbb{N}$ MD transmitters sharing total $n_{\text{tot}}$ subchannels to the same target synchronously. Note that actual subchannel quantity $n_{\text{sub}}^{h,g}=\lfloor \frac{n_{\text{tot}}}{n_{\text{tra}}^{h,g}} \rfloor$ for each task is rounded down. According to \cite{b6}, SINR must surpass upload threshold $\varepsilon_{\text{suc}}$ to accomplish successful transmissions. Thus, success probability $p_{\text{suc}}$ is formulated as:
\begin{align}
\varepsilon_{\text{suc}}^{h,g}=&2^{\frac{S^{h,g}n_{\text{tot}}}{\Delta t W^{h, h^{\prime}}{n_{\text{sub}}^{h,g}}}}-1,\\
p_{\text{suc}}^{h,g}=&e^{\frac{-N_0}{P t^{h,g} (Dt^{h,h^{\prime},g})^{-\alpha}} \varepsilon_{\text{suc}}^{h,g}}.
\end{align}

Suppose that $\kappa$ is the state whether the task is dropped or not. To those failed to be transmitted, their drop state $Ds^{i,g} = \kappa \left( 0,1 \right)$ is defined as 1, illustrating misfit communication source allocation. In terms of those achievers, transmission time $T_{\text{tra}}$, energy cost of transmitter $E_{\text{tra}}$ and receiver's energy cost $E_{\text{rec}}$ are given as:
\begin{align}
T_{\text{tra}}^{h, g} =& \frac{S^{h,g}}{R_{\text{tra}}^{h,g}}, \\
E_{\text{tra}}^{h, g}=&P t^{h,g} T_{\text{tra}}^{h, g}, \\
E_{\text{rec}}^{ h, g}=&P r^{h,g} T_{\text{tra}}^{h,g}.
\end{align}

\textbf{Task Execution \& Failure State:} Due to similar first-in-first-out (FIFO) task processing orders, there is no exact difference between local and remote execution except transmission. 

To figure out task execution time length $T_{\text{exe}}$ for receiver $h$, its completed time point $T_{\text{fin}}$ and beginning one $T_{\text{sta}}$ both should be asseverated in advance, having known task generation time $k$ and $\theta$, which marks that the task is offloaded or executed locally. Notably, we accelerate decided CPU frequency $f$ to $f_{\max}$ once the task runs out of tolerable delay.
\begin{align}
T_{\text{fin}}^{h, g}=&\max \left(T_{\text{fin}}^{h, g-1}+T_{\text{exe }}^{h, g}, k+\theta\left(T_{\text{tra}}^{h, g}\right)+T_{\text{exe }}^{h, g}\right), \\
T_{\text{sta}}^{h, g}=&\max (T_{\text{fin}}^{h, g-1}, k+\theta\left(T_{\text{tra}}^{h, g}\right)),\\
T_{\text{exe}}^{h, g}=&\max 
(0,\min (D^{h,g}+k-T_{\text{sta}}^{h, g}, \frac{C^{h,g}}{f^{h,g}}) ) \nonumber \\
+&\frac{\max \left(0, C^{h,g}-\max (0,\left(D^{h,g}+k-T_{\text{sta}}^{h, g}\right)) f^{h,g}\right)}{f^{h}_{\max }}
\end{align}

In terms of energy consumption, a task is estimated to consume requisite execution energy $E_{\text{exe}}$, comprised of statistic energy to sustain control circuits and additional dynamic energy for task execution \cite{b10}, as well as potential transmission energy $E_{\text{tra}}$. Premise that $K_{\text{stat}}$ means DVFS static power factor while $K_{\text{dyn}}$ is for dynamic one, $E_{\text{exe}}$ is given as:
\begin{align}
E_{\text{exe}}^{h,g}=\left(1+K_{\text{stat}}\right)  K_{\text{dyn}} (V^{h,g})^2  f^{h,g}  T_{\text{exe}}^{h, g}.
\end{align}

Furthermore, as long as the battery depletes energy or the offloading target disconnects, the target's battery level would be observed as 0, leading to equivalent task execution failure. Meanwhile, bounded battery level exposes that if generated task quantity is large enough, the fewer task execution failures, the lower average energy expense of successfully managed ones. Accordingly, to every MD, the failure status of current task $Fs^{i,g}$ can be seen as a key index of decision assessment and is marked as 1 for failures: $Fs^{i,g}=\zeta(0,1)$, where $\zeta$ signifies whether the transmission succeeds or not.

\textbf{Battery Degradation:} We firstly predefine transition point as battery working power $p$, voltage $V$ and discharging current $I$ mutation. Apparently, transmission and operating frequency variation would both experience transition points. With the aid of battery degradation estimated in \cite{b12,b13}, we could recompose it to accommodate offloading scenario. It is assumed that a task is executed from $t_0$ to $t_Z$ with $Z$ transition points and each status $z$ lasts $\Delta t^{z}$ seconds, then SoC average $SoC_{\text{avg}}^g$, standard deviation $SoC_{\text{dev}}^g$, along with effective cycle number $N_{\text{cyc}}^g$ are calculated below to identify battery degradation extent ${BD}^g$ caused by present task:
\begin{align}
&SoC_{\text{avg}}^g 
=\sum_{z=0}^{Z-1}\frac{\left(SoC\left(t^z\right)+SoC\left(t^{z+1}\right)\right) }{2(t^Z-t^0)} \Delta t^z ,\\
&SoC_{\text{dev}}^g=2 \sqrt{\frac{3}{t^Z-t^0} \int_{t^0}^{t^Z}\left(SoC\left(t\right)-S o C_{\text{avg}}^g\right)^2 d t} \nonumber \\
&=2 \sqrt{\sum_{z=0}^{Z-1} \frac{\left(B_{\text{avg}}^g-B\left(t^z\right)+p^z  \Delta t^z\right)^3-\left(B_{\text{avg}}^g-B\left(t^z\right)\right)^3}{p^z \left(t^Z-t^0\right)  B_{\max }^2}},\\
&N_{\text{cyc}}^g =\frac{1}{2 Q_{\text{norm}}} \int_{t^0}^{t^Z}|I(t)| \mathrm{d} t 
=\frac{SoC\left(t^0\right)-SoC\left(t^Z\right)}{2},\\
&BD^g=\left(a N_{\text{cyc}}^g e^{\left(SoC_{\text{dev}}^g-1\right) b}+0.2 \frac{t^Z-t^0}{T_{\text{life}}}\right) c e^{d\left(SoC_{\text{avg}}^g-0.5\right)},
\end{align} 
where $B_{\text{avg}}$ denotes the average battery energy and $Q_{\text{norm}}$ marks the nominal cell capacity. Similar to \cite{b12}, $a, b, c, d$ and $T_{\text{life}}$ are constants related to physical battery features.

\subsection{Problem Formulation}\label{section:A}
According to \cite{b14} for dynamic voltage scaling, one ideal strategy would happen with uniform CPU operating speed and this could be proved the same in DVFS when static execution energy is proportional to the dynamic one. Thus, suddenly accelerating CPU frequency for the out-deadline part would raise energy expenses. Concurrently, network congestion would also boost energy expenditure, since there is less time left for execution, while higher CPU operating speed innately rockets energy spending. Therefore, energy consumption is an efficacious criterion for delay satisfaction and decision-making. Consequently, to possess a better mastery of user preference and energy expense synchronously, an expedient variable, prioritized energy $E_{\text{pri}}$, is introduced as:
\begin{align}
&E_{\text{pri}}(h, g, h^{\prime}, g^{\prime} )=X^{h , g}\left(E_{\text{tra}}^{h, g}+E_{\text{rec}}^{h, g}+E_{\text{exe}}^{h^{\prime}, g^{\prime}}\right)  \nonumber \\
&=\left\{\begin{array}{l}X^{h, g} E_{\text{exe }}^{h^{\prime}, g^{\prime}}, \text{ if } h^{\prime}=h;  \\
X^{h, g}\left(E_{\text{tra}}^{h, g}+E_{\text{rec}}^{h, g}+E_{\text{exe}}^{h^{\prime}, g^{\prime}}\right), \text{ if } h^{\prime} \neq h. 
\end{array}\right.
\end{align}

Note that $h, g$ tag $g$th task produced by MD $h$ which is $g^{\prime}$th task executed by target $h^{\prime}$. This is effectual for that with equal $E_{\text{pri}}$, task owning a relatively small $X$ is affirmed to have more energy for speedy execution. On the other hand, ascending energy spent per successfully managed task is speculated to reflect more task failure and irrational network jams. 

Energy and communication resource allocation considering user preference, network status, SoH could be generally governed by four variables: prioritized energy, failure state, drop state and battery degradation. If we predefine $BD_{tot}^g$ as the battery degradation level of both transmitter and receiver, then
\begin{equation}
B D_{\text{tot}}\left(h, g, h^{\prime}, g^{\prime}\right)=B D^{h,g}+B D^{h^{\prime},{g^{\prime}}} .
\end{equation}

Thanks to extrapolated execution order $g^\prime$ in target $h^\prime$ having known the production order $g$ in transmitter $h$, or in other words, task generated MD $i$, the cost of every task can be estimated as the following:
\begin{align} 
&cos t\left(i, g \right) = cos t\left(h, g,  h^{\prime},g^{\prime}\right)\label{1}\\
&=\omega E_{\text{pri}}\left(i,g\right)+\beta Ds(i, g) +\gamma Fs(i, g) +\chi BD_{\text{tot}}\left(i,g\right) \nonumber.
\end{align}
where $\omega,\beta,\gamma,\chi$ are constants catering for contingent user desire. Furthermore,  for every agent, its objective is to minimize the individual cost sum of generated tasks through manipulating offloading decision $\mathcal{O}$ and CPU operating frequency $\mathcal{F}$. Thence, the optimization problem is defined as:
\begin{align}
\textbf{P}:\min_{\mathcal{O},\mathcal{F}} & \sum_{g=1}^{G_{\text{tot}}^i} cost(i, g), \forall i \in \mathbb{N}  \\
\text{ s.t. }  \sum_{g=1}^{G_{\text{tot}}^h} E_{\text{tra}}^{\text{h,g }}+&\sum_{g=1}^{G_{\text{pro}} ^h} \left(E_{\text{exe}}^{h, g}+E_{\text{rec}}^{h, g}\right) \leqslant B_{\text{max}}, \forall h \in Q 
\label{2},
\end{align}
where $G_{\text{tot}}$ and $G_{\text{pro}}$ are the total task quantity generated and processed by each device. Constraint \eqref{2} indicates that total energy spending must be within battery limitation $B_{\text{max}}$. In the following part, we convert such a problem into an MAMDP for solution exploration in the dynamic scene.

\section{Problem Solution}
The MAMDP is written as: 
$\mathcal{M} = \left(\mathcal{S} ,\mathcal{A},\mathcal{P},\mathcal{R}\right)$. Here, $\mathcal{S}$ and $\mathcal{A}$ represent state space and action space respectively. $\mathcal{P}\left(s^{\prime} \mid s, A\right)$ denotes transition possibility from $s$ to $s^{\prime}$ with individual reward $r^i \in \mathcal{R}$ gained by each agent after taking joint action $A=\left\{a^i\right\}_{i=1}^M$, where every agent's action is $a^i=\left(\mathcal{O}^i, \mathcal{F}^i\right)$ .

\subsection{State and Action Space}
To obtain a complete observation of channel status, device state and task property, state space $\mathcal{S}$ is defined as: 
\begin{align}
\mathcal{S}\triangleq \{s=(&\left\{D t^i\right\}_{i=1}^M,\left\{S^i\right\}_{i=1}^M,\left\{B^h\right\}_{h=1}^{M+N}, \nonumber \\
&\left\{T_{\text{fin}}^h\right\}_{h=1}^{M+N},\left\{C^i\right\}_{i=1}^M,\left\{X^i\right\}_{i=1}^M) \}    
\end{align}
where $Dt$ and $S$ affect transmission success possibility and rate, battery level $B$ impacts executable workload, $T_{\text{fin}}$ cares deadline satisfaction, $C$ is associated with execution time consumption, $X$ indicates subscriber preference. We additionally set action space with $WF$ CPU operating modes as:
\begin{align}
\mathcal{A}\triangleq\{A=\{(\mathcal{O},\mathcal{F})\}_{i=1}^M\mid &\mathcal{O} \in [1, M+N], \nonumber \\&\mathcal{F} \in [1,W F]\} 
\end{align}

Cautiously, every agent needs to be responsible for its own action options and rewards but cooperate together to optimize the overall offloading schedule.

\subsection{Reward}
Having discussed in Section \ref{section:A}, all energy consumption could be regulated by prioritized energy and dropped task number concerning execution deadline fulfillment; network congestion could be curbed by transmission failure state; and SoH preservation is directly controlled by battery degradation extend. Hence, the reward function after adopting action $A$ in state $s$ is computed as
\begin{align}
&r^i(s(t), A(t))=-\tau cost(i,t)=-\tau cost(i,g)\\
&=-\tau(\omega E_{\text{pri}}\left(i,g\right)+\beta Ds(i, g) +\gamma Fs(i, g) +\chi BD_{\text{tot}}\left(i,g\right) ) \nonumber
\end{align}

Since the generated order and time of a task are one-to-one corresponding, $cost(i,g)$ can be rewritten as $cost(i,t)$. Because DRL aspires to maximize cumulative rewards, rewards are designed as minus ones to realize the minimization in $\textbf{P}$. A constant $\tau$ is multiplied for visual convenience. Thus, as sets of actions are taken with time passing by, individual cumulative reward $R^i = \sum_{t=1}^T r^i(s(t), A(t))$ is maximized. Notably, the maximizing problem is equivalent to the original \textbf{P}.

\subsection{Algorithm}
We intend to address the problem in a multi-agent way, especially in a MCC manner with non-continuous actions. Consequently, we propose a discrete rMAPPO method with separate rewards to sort out the problem, which enjoys excellent performance. 

Each agent would experience the same training process as proximal policy optimization (PPO), possessing actor and critical networks. Here, we configure networks as RNNs to accumulate loss functions while utilizing backpropagation through time (BPTT) to train the network. Although similar to PPO, rMAPPO would undergo centralized training and decentralized execution. The core difference between it and independent PPO is the global value function input for rMAPPO while a local one for the latter. Here, global input combines duplication-eliminated local observation with global environmental information, forming a fully observed MAMDP. This input assists the critic network in calculating the centralized Q function, facilitating the actor network to update parameters and determine actions.
\begin{algorithm}
\caption{rMAPPO and DVFS-powered task   offloading solution with separate reward } \label{alg2}
    \SetAlgoLined
    \SetKwInOut{Input}{Input}
    \SetKwInOut{Output}{Output}
    \Input{The maximum total steps $step_{\max}$, maximum time step per episode $T$, agent set $m \in \mathbb{M} $, batch size $\mathcal{B}$ and mini-batch size $\mathcal{K}$}
    Initialize actor network $\pi$ and critic network $V$ with parameters $\theta$,$\phi $ respectively\\
    \While{$step \leq step_{\max}$ }
    {Initialize data buffer $\mathcal{D}$ as an empty directory \\
      \For{$i=1$ to $\mathcal{B}$}
      { Set an empty list $\tau $\\
        Initialize actor RNN states $h_{\pi}^{0 ,m} $ and  critic RNN states $h_{V}^{0,m}$ for each agent $m$\\
      \For{$ t=1$ to $T$} {
      \For{each agent $m$}
      {Produce current actor state $h^{t,m}_{\pi}$, action set of working frequency and offloading target $a^{t,m} = (\mathcal{O}^{t,m},\mathcal{F}^{t,m})$ from actor network\\
      Produce current critic state $h^{t,m}_V$ from critic network
      }
      Interact with environment based on joint actions $A^t$, then obtain rewards $r^{t,m}$ for each agent $m$, next states $s^{t+1}$, next observation $o^{t+1}$\\
      Store $[s^{t},o^{t},h^{t}_{\pi },h^{t}_V,A^{t},r^{t},s^{t+1},o^{t+1}]$ into the list $\tau $
      }
      Apply PopArt to calculate advantage estimate $\hat{A}$ along with reward-to-go $\hat{R}$\\
      Separate $\tau $ into chunks then reserve chunks into data buffer $\mathcal{D}$\\
      }
      \For{each mini-batch $k = 1,\ldots,\mathcal{K}$}
      {Randomly choose mini-batch $d$ from $\mathcal{D}$ \\
      Repeat update RNN hidden states for $\pi,V$ using the first hidden states in every chunk within $d$
      }
      Update $\theta $ with and $\phi $ using Adam
     }
\end{algorithm}

The overview of the proposed rMAPPO algorithm is shown in Algorithm~\ref{alg2}, accompanying two essences: exploration and update stages. For exploration, shown in lines 7-14, after initializing actor and critic state, in slots within episode length $T$, rMAPPO should decide the offloading targets along with the CPU operating frequency of these targets, and interacts with the environment before agents receive their own rewards. Subsequently, local observation and global environmental information are collected and summarized. Until now, the advantage estimate could be calculated by generalized advantage estimation. 
All information would then be diced and stored in the buffer. 

The update stage arises in lines 18-23. Aiming at stable training, in batch-sized epochs, rMAPPO randomly samples mini-batch data within the data buffer and harnesses data chunks to iterate the network. The intensified network helps to pursue distinguished goals of networks, namely maximizing object function for actor network and minimizing loss function for critic network. 
\section{Simulation Result}
In this section, simulations are conducted to verify the effectiveness of the proposed offloading scheme. Note that, for a downward trend, we negate the costs. We suppose that there is a $\SI{200}{\metre} \times \SI{200}{\metre}$ room with $N = 7$ randomly distributed MDs and a base station located at the center. The base station is connected with $M = 5$ EDs and covers the complete room, whereas an upper limit of $\SI{30}{\metre}$ to the D2D link. Assuming equalized transmitting and receiving power, transmitting power is designed as $Pt_{\text{m}} = 100\mathrm{~mW}$ and $Pt_{\text{e}} = 200\mathrm{~mW}$. Besides identical bandwidth $W = 10\mathrm{MHz}$ to each receiver provided by $n_{\text{tot}}= 64$ subchannels, path loss exponent $\alpha = 4$ and noise $N_0 = 5\times10^{-14}\mathrm{W}$ are also assumed. With consideration to devices, we presuppose that ED and MD own the same battery capacity $B_{\text{max}} = 1000\mathrm{J}$ while holding different DVFS configurations. Although the static power factor behaves alike, equaling $0.3$ \cite{b10}, MDs utilize AMD Turion MT-34 single-core processors, and EDs harness AMD Opteron 2218 dual-core ones \cite{b15}. We posit that dual cores would handle the same task with frequency adjusting granularity set as whole equipment. Additionally, device disconnection follows a binomial distribution $B(T,0.1)$, and tasks are produced by a possibility of $0.9$ in each slot at every MD. The battery degradation arguments: $a,b,c,d,T_{\text{life}}$ are similar to \cite{b12}. Finally, the other parameters in the simulation are listed in Table \ref{tab:Parameters}. Notice that due to the significance of energy saving and the unavoidably worse delay after disasters\cite{b2}, we utilize the clipped-delay approach  as \cite{b14} in simulation with deadline $D_{\text{limit }}$ to each task category for more delay tolerance.
\begin{table}[htb]
\centering
\vspace{-0.3cm}
\caption{TASK PARAMETERS. }
\begin{tabular}{ll}
\hline Parameters & \multicolumn{1}{c}{ Values } \\
\hline$S^g$ &\cellcolor{mygray} $S \sim U(5,6) \mathrm{Mbit}$ \\
\hline$C^g$ &\cellcolor{mygreen} $C \sim U(0.2,0.6) \mathrm{GHz}$ \\
\hline \multirow{2}{*}{$D^g$} & \cellcolor{mygray} $D \sim U(0.05,0.5) \mathrm{s}, D_{\text{limit }}=0.3$ if $X_g=1$ \\
&\cellcolor{mygreen} $D \sim U(0.5,1) \mathrm{s}, D_{\text{limit }}=0.75$ if $X_g=2$ \\
& \cellcolor{mygray} $D \sim U(1,3) \mathrm{s}, D_{\text{limit }}=2$ if $X_g=3$ \\
\hline
\end{tabular}
\label{tab:Parameters}
\end{table}

\begin{figure}[htb]
\vspace{-0.7cm}
    \centering
    \includegraphics[width=0.7\linewidth]{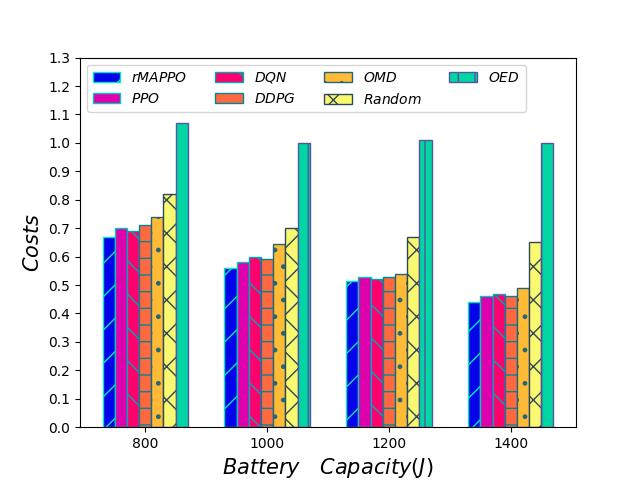}   
    \caption{Costs with different battery energy limits.}
    \label{3}
\end{figure}
\begin{figure}[htb]
\vspace{-0.7cm}
    \centering
    \includegraphics[width=0.7\linewidth]{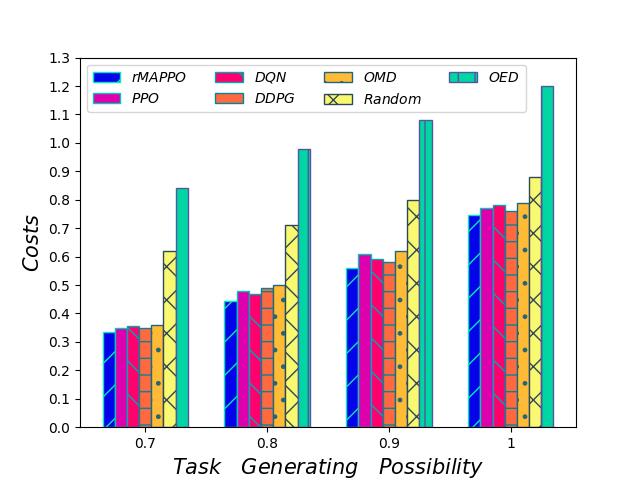}   
    \caption{Costs under various task generating probabilities.}
    \label{4}
\end{figure}

Fig.~\ref{3} and Fig.~\ref{4} present the performance of rMAPPO and other baseline approaches under various environments. Aside from the shrinking total costs as battery capacity enlarged or task quantity dwindled, it is apparent that rMAPPO always possesses the best outcome among the competition with following traditional baseline algorithms:\\
$\bullet$ PPO: A mainstream policy-gradient-based DRL approach. \\
$\bullet$ DDPG (deep deterministic policy gradient): A baseline algorithm compatible with policy learning and Q-value learning. \\
$\bullet$ DQN (deep Q-network): A conventional DRL method coupling Q-learning with deep neural network.\\
$\bullet$ OED (offloading to EDs): Offload all tasks to EDs. \\
$\bullet$ OMD (offloading to MDs): Execute all tasks in MDs.\\
$\bullet$ Random: Randomly choose offloading targets and CPU operating frequencies.

Despite traditional baseline approaches, it is necessary to compare the ability of rMAPPO with the representative MADRL method, taking multi-agent DDPG (MADDPG) for example. Fig.~\ref{4} illustrates that rMAPPO finally outperforms MADDPG. This might be because of Q-value overestimation, which occurs less frequently in rMAPPO \cite{b16}.
\begin{figure}[htb]
\vspace{-0.45cm}
    \centering
    \includegraphics[width=0.7\linewidth]{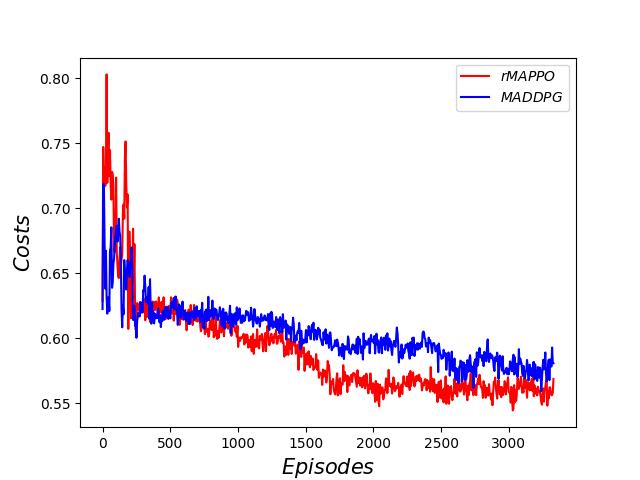}   
    \caption{Convergence tendencies of multi-agent algorithms.}
    \label{5}
\end{figure}

Additionally, since DVFS is adopted, it is necessary to evaluate the outcome of DVFS by competition with uniform operating frequency. It could be ascertained that DVFS actually balances the requirement between delay and energy consumption in Fig.~\ref{5}, certifying its potency.
\begin{figure}[htb]
\vspace{-0.42cm}
    \centering
    \includegraphics[width=0.7\linewidth]{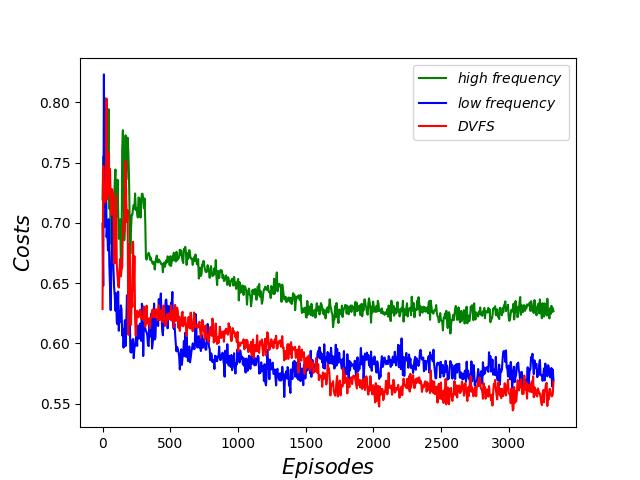}   
    \caption{Performances under different CPU speed mechanisms.}
    \label{6}
\end{figure}

\section{Conclusion}
In this paper, we consider a D2D-supported post-disaster task offloading scene within an unstable network. Further concerns about user preference as well as battery SoH are also formally discussed. The resource allocation and offloading decision problem is handled through a DVFS-aided rMAPPO method with separate rewards due to the lack of a centralized controller and requirement trade-off. After comparing with conventional baseline algorithms and representative multi-agent approach MADDPG, our offloading strategy exhibits its effectiveness in convergence results. We also demonstrate the assistance of DVFS against fixed-speed execution, counterbalancing all decision-impacting factors.

\bibliographystyle{IEEEtran}
\normalem
\bibliography{ref}

\end{document}